# Introduction to the Mechanical Design of Accelerators

*M. Timmins, CERN, Geneva, Switzerland*

**Abstract**

Particle accelerators represent some of the most sophisticated engineering achievements of our time. Their construction requires a unique combination of physics insight and mechanical engineering expertise. The aim of this paper is to provide young mechanical engineers with an introduction to the principles, methods, and challenges associated with the mechanical design of accelerators. The lecture upon which this proceeding is based emphasized the translation of functional requirements into engineering specifications, the critical importance of robust and reliable design, and the need for precisely defined drawings supported by international standards such as ISO GPS [2] and GD&T [3]. Through illustrative examples and a practical case inspired by CERN's existing components, the paper underlines the necessity of anticipating lifecycle demands, ensuring manufacturability, and safeguarding operational reliability. Particular emphasis is placed on the contractual value of 2D drawings, the practical application of functional dimensioning, and tolerance chain analysis. By reviewing common pitfalls and exploring best practices, the paper seeks to orient engineers towards design choices that balance cost, reliability, and performance.

## 1    Introduction

Particle accelerators are indispensable tools for modern physics. They produce beams that enable groundbreaking discoveries in particle physics, nuclear research, and materials science. Given the societal investment in such infrastructures—the LHC alone representing one of the most costly scientific projects ever—mechanical reliability is of paramount importance. Downtime can be extremely expensive; for example, a single day without beam in the LHC costs approximately 200,000 CHF. This economic weight underscores the vital role of robust design. Mechanical engineers play a central role in ensuring accelerator components remain operational for decades, often under harsh conditions of radiation, thermal stresses, and precise alignment requirements. Examples of past failures, such as damaged bellows, RF finger failures, and ripped jacks in the LHC, as shown in Fig. 1, illustrate the risks of insufficiently robust designs. Such cases highlight the need for thorough translation of functional requirements into measurable, enforceable specifications.

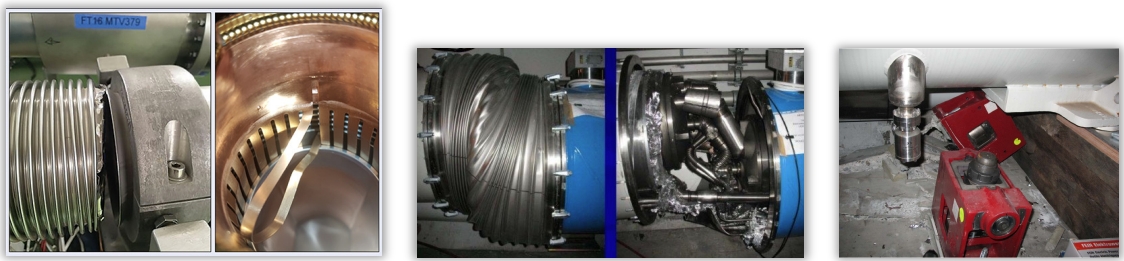

**Fig. 1:** Illustrations of failures in the LHC

## 2   Functional Requirements and Mechanical Specifications

The foundation of accelerator design lies in accurately identifying functional requirements. These requirements must then be translated into engineering specifications—parameters such as dimensional tolerances, materials, joining techniques, coatings, and assembly procedures. Achieving this translation is challenging because specifications must be both reachable and measurable. The lifecycle of accelerator components as shown in Fig. 2 must also be considered, covering construction, assembly, installation, maintenance, future upgrades and decommissioning stage including dismantling, and waste management. Each stage imposes distinct constraints, from ease of alignment during installation to repeated handling operations in harsh environments. Two broad steps structure the design approach: preliminary conceptual design and detailed design. The first aims to satisfy most critical functional requirements at a higher level through conceptual material choices, preliminary dimensions, cooling considerations if needed, and cost estimates, accompanied by first-pass thermal and structural analyses. The second involves exhaustive 3D models and, critically, 2D specification drawings, which form the contractual basis for fabrication. Reliance solely on 3D models risks overlooking essential specifications such as coatings, tolerances, and weld details.

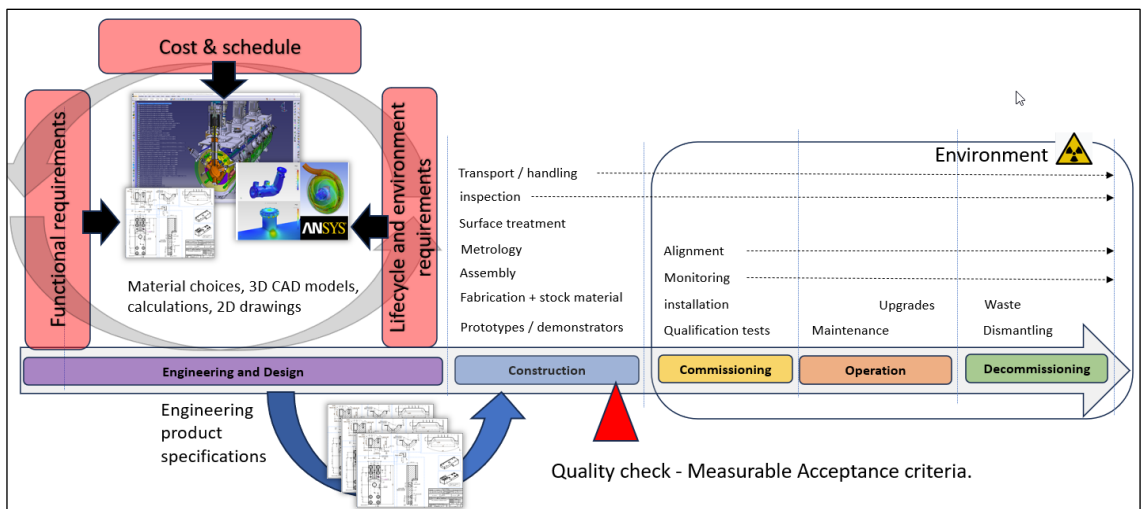

**Fig. 2:** Product lifecycle of accelerator components

## 3   Best Practices in Mechanical Design

Best practices encourage engineers to start from proven concepts that satisfy a majority of requirements before venturing into novel solutions. Designing to established norms—such as pressure vessel code EN13445[4], design of steel structures Eurocode 3 [5], standard bolted connections VDI 2230 [6], and handbook practices—anchors the design in tested knowledge and reduces risk. More specifically, as commonly used in designs of accelerator components, welding or brazing presents specific challenges: local heating can lead to large deformations, compromised precision, and degraded mechanical properties, particularly in materials such as aluminum and copper. Leaks and inspection difficulties further complicate welded assemblies, for this reason, "the best weld is no weld" is a guiding principle in accelerator design. Maintenance considerations are equally critical; systems designed for easy disassembly using sealed bolted joints may introduce weaknesses if not carefully managed. Thus, design of a mechanical joint is always a trade-off between functionality, cost, and reliability as shown in Fig. 3.

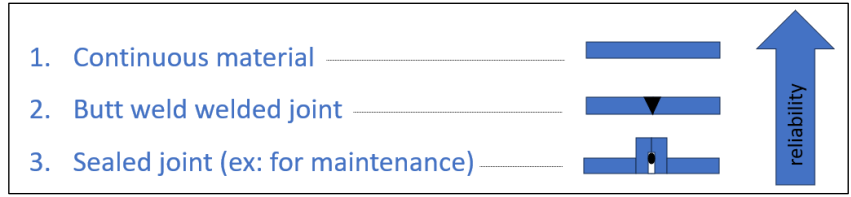

**Fig. 3:** Reliability for mechanical joint

As outlined in the introduction, bellows (or compensators), are extensively employed in accelerator systems to guarantee vacuum continuity, yet their inherently thin-walled construction, (typically 200 microns) often renders them a critical point of failure. The two predominant types available on the market, illustrated in Fig. 4, display markedly different reliability profiles. Hydroformed bellows, produced by shaping a continuous tube into convolutions, generally exhibit superior structural integrity. In contrast, edge-welded bellows are composed of multiple convolutions joined by welds, which not only increases the risk of leakage at the numerous weld seams but also creates interstitial regions where contaminants may accumulate. Such trapped pollution can promote localized pitting corrosion, further undermining long-term reliability. Consequently, the selection of an appropriate bellows configuration must be informed by a rigorous requirements analysis, ensuring a carefully optimized balance between mechanical flexibility and operational robustness.

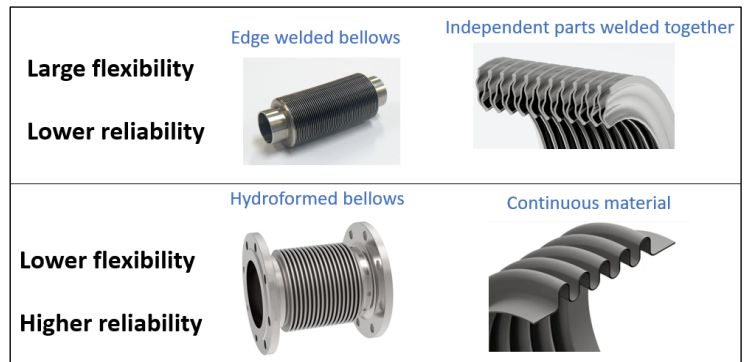

**Fig. 4:** Edge welded bellows vs Hydroformed bellows

## 4    Importance of 2D Drawings and ISO GPS

Despite advances in digital modelling, 2D drawings remain the authoritative contractual documents for manufacturing accelerator components. These drawings encapsulate the complete set of requirements, including dimensions, tolerances, coatings, and assembly instructions. Relying solely on 3D models for subcontracted work can lead to dangerous misinterpretations, which is why it is essential that the same team—or closely collaborating engineers—oversee both models and drawings.

The strength of robust 2D drawings lies in functional dimensioning and the adoption of international standards such as ISO GPS (Geometrical Product Specification). Unlike traditional dimensioning, illustrated in Fig. 5, which frequently generates ambiguities, functional dimensioning explicitly links specifications to the intended function, as shown in Fig. 6 through functions named as F01, F02 and F03.

As illustrated in Fig. 6, central to this approach is the correct use of datum systems. A well-chosen datum system systematically constrains all six degrees of freedom—three translational (X, Y, Z) and three rotational (about X, Y, Z)—in a controlled sequence:

The **primary datum** is defined by the most critical functional surface. It typically establishes a plane that removes three degrees of freedom (one translation and two rotations), fixing the part in its fundamental orientation.

The **secondary datum** reflects the next most important functional interface, usually constraining two further degrees of freedom (one translation and one rotation) by aligning the part relative to an edge, axis, or surface.

The **tertiary datum** resolves the last remaining degree of freedom, which may be either a translation or a rotation depending on the part's geometry and the way it is positioned within the assembly. This is often achieved through a point, hole, or slot, providing a definitive and unambiguous location.

The summary of such a chosen datum system and its role in constraining the six degrees of freedom can be conveniently displayed in an isostatic table, as illustrated in Fig. 6. This structured representation reinforces clarity for both design and verification.

By defining primary, secondary, and tertiary datums in this order—always guided by function and assembly conditions—the drawing ensures that the part is positioned consistently with its intended role. This structured constraint of six degrees of freedom not only provides clarity to manufacturers and metrologists but also reduces ambiguity, enhances alignment between design and fabrication, and optimizes costs. The ISO GPS framework thus establishes a precise and shared technical language, strengthening collaboration between designers, fabricators, and quality controllers.

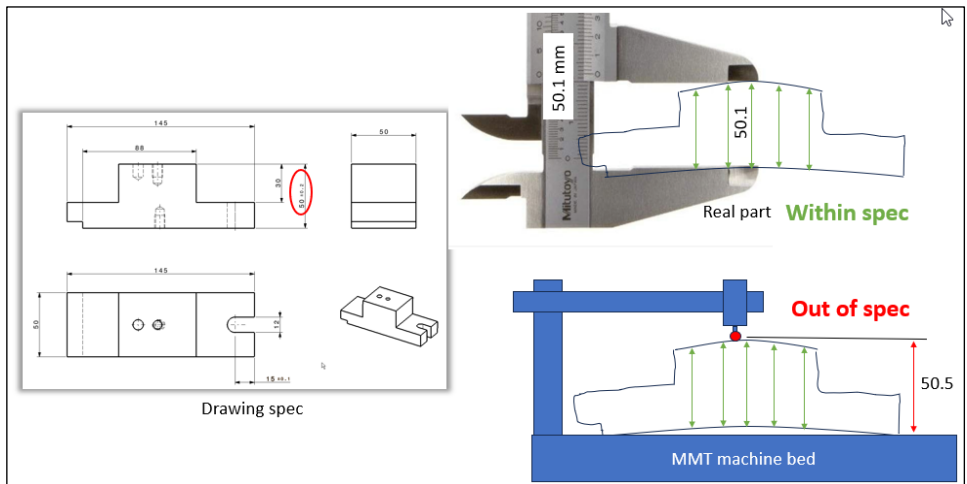

**Fig. 5:** Ambiguities in traditional tolerancing

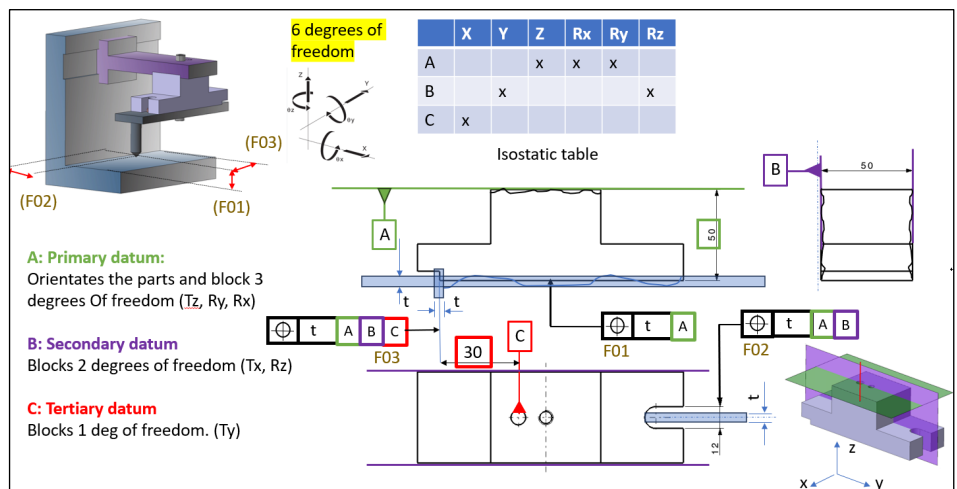

**Fig. 6:** Functional dimensioning using datum features

# 5 Practical Case Study: Scraper Design

To illustrate these principles, the lecture presented a practical case, as shown in Fig. 7, inspired by a scraper design for CERN's SPS machine. The scraper assembly involves multiple components—blade, blade support, vacuum vessel, shafts, and locating pins—whose alignment must meet strict tolerances. Functional tolerancing chains were developed for two specifications, F01 and F02, as shown in Fig. 7 corresponding to critical alignment distances. Analysis using the sum-of-tolerances method revealed in Fig. 8 that F01 fell outside specifications ($10 \pm 0.61$ mm versus required $10 \pm 0.5$ mm), while F02 was achievable ($10 \pm 0.32$ mm). Potential approaches to address out-of-spec results, as shown in Table 1, include reducing the number of parts, tightening tolerances, or adding adjustment mechanisms. Each solution affects both cost and system robustness differently: reducing component count generally provides the most cost-effective option, while maintaining robustness and, consequently, reliability.

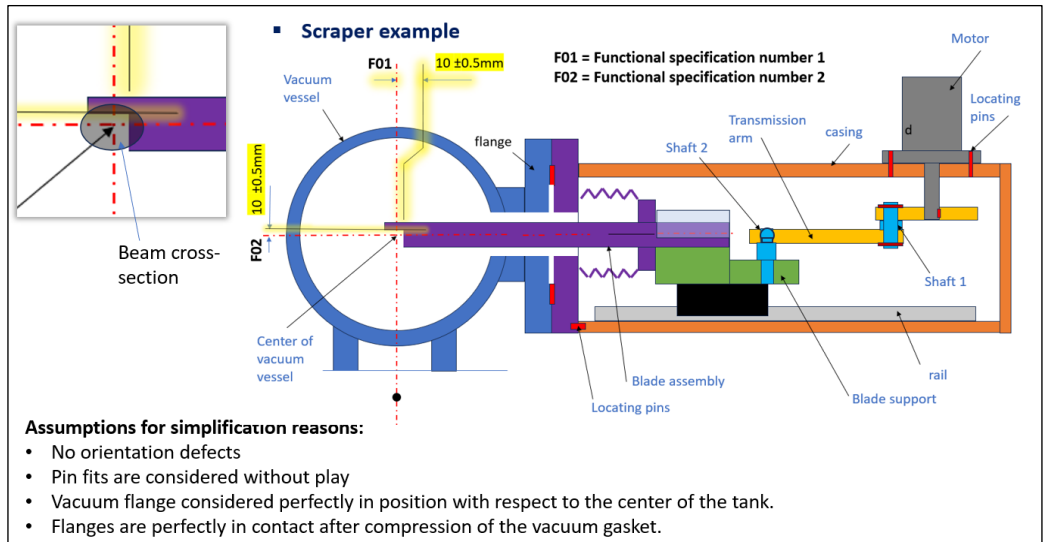

**Fig. 7:** SPS Scrapper design

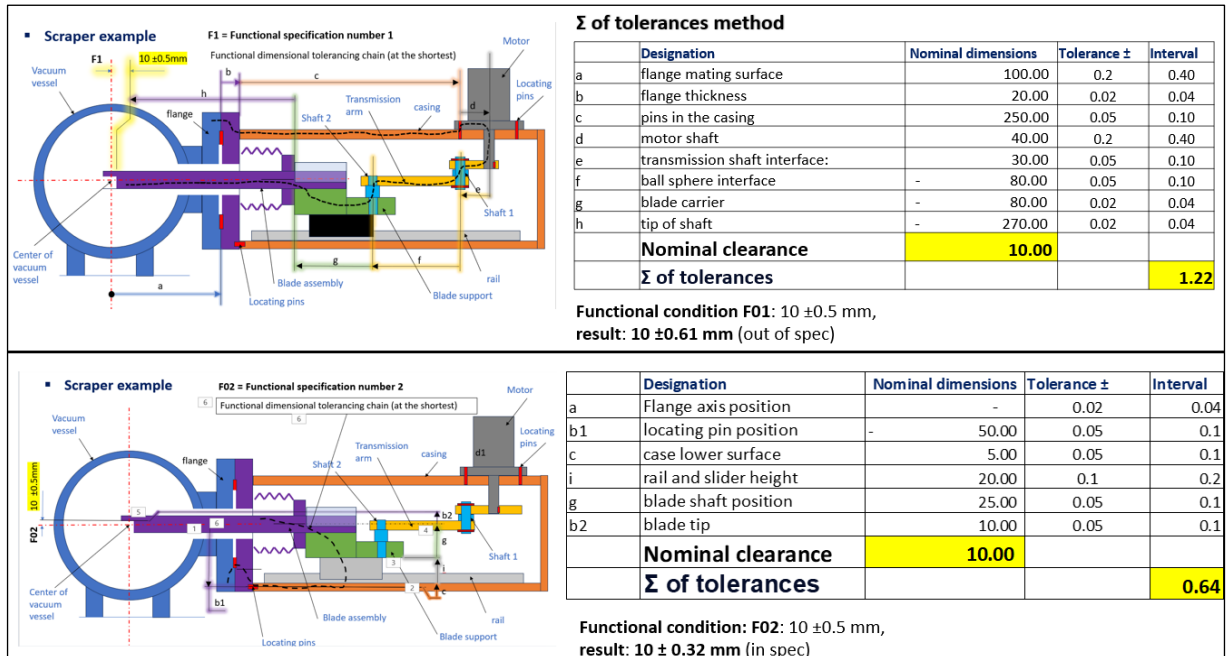

**Fig. 8:** Tolerance stacking calculations

**Table 1:** Solutions to fulfil F01

| Possible solution | Impact on cost | Impact on robustness |
|---|---|---|
| Reduce part count | medium | Low |
| Tighten tolerances | High | medium |
| Add adjustment system | medium | High |

In continuation with the same case study, attention is now directed to the significance of employing ISO GPS dimensioning and tolerancing in order to accurately define the functional dimensional requirements of a system component. The blade support has been selected to exemplify this approach. As discussed in Chapter 4, it is necessary to establish a reference system constraining all six degrees of freedom so as to replicate the actual assembly conditions. This procedure simultaneously enhances clarity in both metrological assessment and manufacturing processes.

As show in Fig. 9, the primary, secondary, and tertiary datum features were determined with particular care to ensure consistency with the assembly constraints. Subsequently, the dimensions influencing specifications F01 and F02, boxed in green in Fig. 9, were identified and assigned appropriate tolerance values through the ISO GPS position symbol, thereby providing an unambiguous representation. Dimensions not directly related to the functional requirements F01 and F02 were instead specified using conventional dimensioning practices, with tolerances defined according to ISO 2768 [1], primarily governed by the capabilities of the manufacturing equipment.

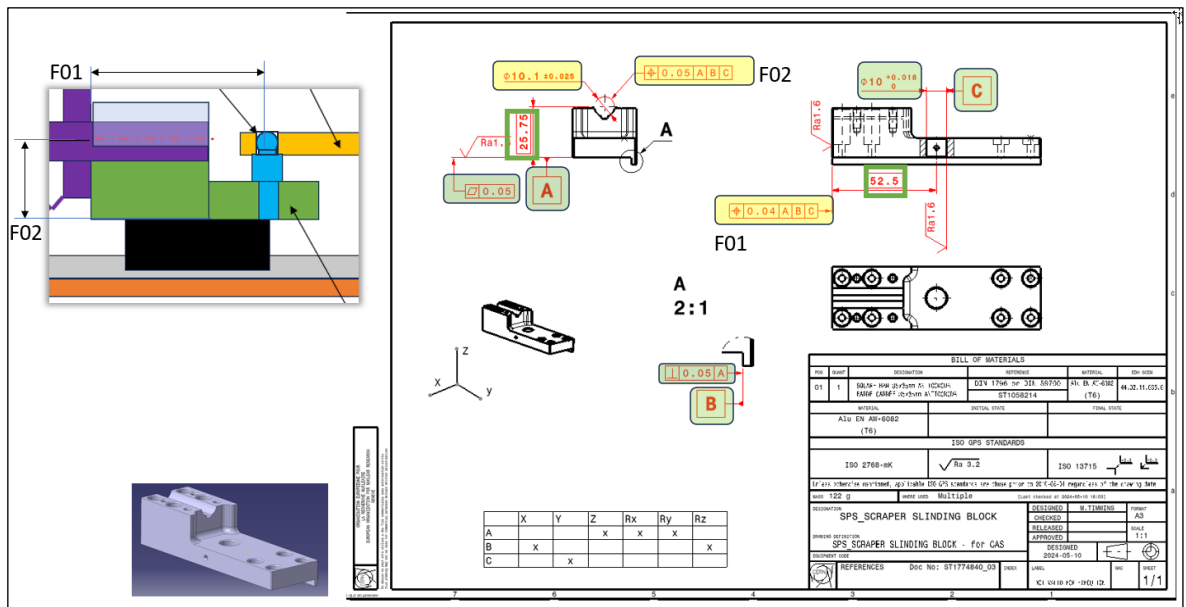

**Fig. 9:** Datum system and ISO GPS functional dimensioning of the blade support

To enhance the overall clarity of the specification, it may be advantageous to establish a distinction between functional dimensions, defined in accordance with ISO GPS notation, and non-functional dimensions, specified using conventional methods, through the application of a color-coding system, as exemplified in Fig. 10.

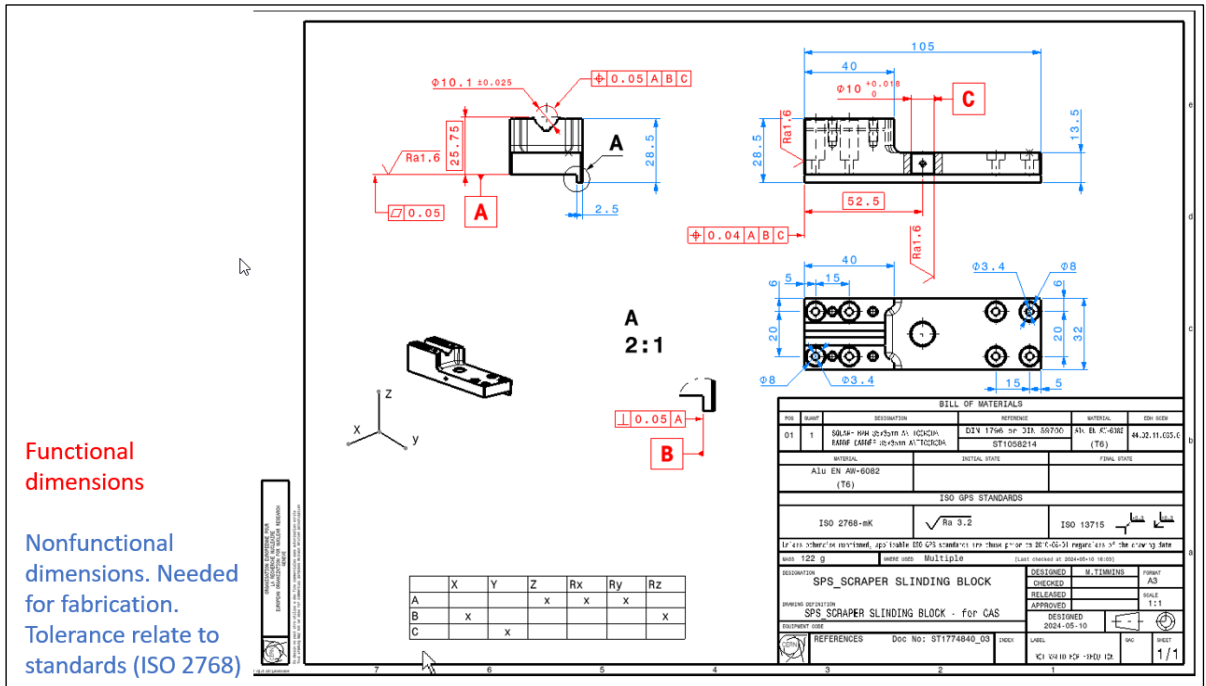

**Fig. 10:** ISO GPS notation and non-functional dimensions expressed through a colour code

This use case demonstrates the importance of balancing precision, manufacturability, and cost. It also highlights how functional dimensioning and tolerance analysis directly guide design choices in high-stakes applications such as accelerators.

# 6    Conclusion

The mechanical design of accelerators demands a synthesis of rigorous engineering, attention to lifecycle, and unwavering focus on reliability. For young engineers, the key lessons include: (1) robust designs minimize costly machine downtime, (2) lifecycle and environmental requirements must be fully integrated into design thinking, (3) 2D drawings remain the binding contractual specifications, and (4) functional dimensioning via ISO GPS provides clarity and reduces ambiguities—an approach reinforced by practical case studies that demonstrate its effectiveness in balancing cost, reliability, and precision. Looking forward, the culture of careful, standards-based design will remain fundamental as accelerators continue to evolve. Equally essential is the recognition that engineering specifications must always be both *reachable* and *measurable*, since only requirements that can be clearly defined, verified, and achieved in practice will translate into reliable and maintainable systems. For mechanical engineers entering this field, embracing these principles ensures they can contribute to machines that not only advance physics but also stand as enduring achievements of engineering.

**Acknowledgement**

— **Company CETISO (J.Y Jacotin):** For the training on ISO GPS and GD& T applications based on industrial practices.

— **Groupe d'Aide à une Cotation de Qualité GRACQ (T. Capelli) :** For reviewing the lecture content.

— **CERN EN/MME design office member:** For their valuable input and exchanges.